# High Thermal Conductivity in Back-End-of-Line Compatible AlN Thin Films

Xufei Guo[1,2,#], Zirou Chen[1,#], Zifeng Huang[1], Yuxiang Wang[1], Jinwen Liu[1,3], Zhe Cheng[1,4,*]

[1] School of Integrated Circuits, Peking University, Beijing 100871, China

[2] School of Electronics Engineering and Computer Science, Peking University, Beijing 100871, China

[3] School of Software and Microelectronics, Peking University, Beijing 100871, China

[4] Frontiers Science Center for Nano-optoelectronics, Peking University, Beijing 100871, China

[#] These authors contributed equally.

*Author to whom correspondence should be addressed: zhe.cheng@pku.edu.cn

**Keywords**: aluminum nitride, thin film, high-thermal-conductivity material, back-end-of-line compatible, time-domain thermoreflectance

**Abstract**

With thermal issues becoming a major challenge to the development of integrated circuits (ICs), high-thermal-conductivity (high-TC) materials are gaining interest from both the industry and academia, especially for high-density back-end-of-line (BEOL) structures. Aluminum nitride (AlN) is an insulating material with high TC, suitable for thermal management in electronic devices. Furthermore, polycrystalline AlN thin films can be deposited at BEOL-compatible low temperatures (< 400 °C) while retaining relatively high TC, rendering AlN a promising candidate for BEOL heat dissipation. Still, AlN deposition should aim at achieving high TC on a variety of substrates used in IC processes. In this work, we obtained 600- and 1200-nm BEOL-compatible AlN thin film samples on Si<111>, SiO, SiN, and $Al_2O_3$ substrates for structural and thermal characterizations. Specifically, time-domain thermoreflectance (TDTR) measurements revealed consistently high TC (> 45 W $m^{-1}$ $K^{-1}$) on different substrates. Moreover, finite element analysis (FEA) simulations of AlN capped on top of an indium-tin-oxide (ITO) transistor showed a reduction of up to 44% in peak device temperature. Our work provided experimental and calculational evidence for the practicality of leveraging AlN as a high-TC, BEOL-compatible heat spreader material.

## 1. Introduction

With the rise of artificial intelligence (AI) and autonomous systems, transistor dimensions have been continuously scaled to meet the increasing demand for high-performance computing (HPC)[1]. However, as the integration and power density of transistors continue to increase, self-heating effects (SHE) have become a severe thermal challenge to device performance and reliability[2], [3], [4]. In advanced logic devices and high-power applications, such as GAAFETs and AlGaN/GaN HEMTs, localized hotspots with extremely high temperatures can reduce carrier mobility and increase leakage, thereby accelerating device degradation[5], [6]. This thermal challenge is particularly acute in systems with complex back-end-of-line (BEOL) structures, as heat generated in the active devices must be dissipated through a stack of intricate interconnects and dielectric layers typically with low thermal conductivity (TC)[7]. Consequently, BEOL-compatible dielectric materials with higher TC is highly desirable to provide efficient lateral and vertical heat dissipation paths[8].

Among dielectric materials, aluminum nitride (AlN) has attracted significant attention due to its ultrawide bandgap (~6.2 eV) and high intrinsic TC (321 W m$^{-1}$ K$^{-1}$) in bulk crystals[9], [10], [11]. However, BEOL-compatible integration requires AlN to be deposited at low temperature (< 400 °C) to avoid degradation of the underlying devices[12], [13]. Such temperature constraints typically result in high densities of grain boundaries and defects, which scatter phonons and reduce the TC well below bulk values[14], [15]. Additionally, as AlN films are thinned to sub-micrometer scale, phonon-boundary scattering strongly suppresses thermal conductivity, which further deteriorates heat transfer capability[16],

[17], [18].

Significant progress has been made in the BEOL-compatible integration and thermal transport characterization of AlN thin films. These studies have demonstrated that the thermal conductivity of AlN thin films is strongly affected by film thickness, defect density and crystalline quality induced by low growth temperatures[18], [19], [20], [21], [22], [23]. Several studies have examined the influence of substrate properties, such as grain orientation and surface condition, on the thermal behavior of AlN thin films[22]. However, existing work focused on only one or two substrate types, most commonly crystalline silicon or sapphire[19]. Furthermore, systematic research integrating experimental thermal characterizations of AlN thin films with simulations of practical applications remains limited.

In this work, AlN thin films with thicknesses of 600 nm and 1200 nm were deposited via magnetron sputtering at a BEOL-compatible temperature of 400 °C. A variety of substrates were considered, including Si <111>, SiO, SiN, and $Al_2O_3$ (sapphire). The microstructure and crystalline quality of the films were characterized using scanning transmission electron microscopy (STEM) and X-ray diffraction (XRD). Thermal conductivity was measured by time-domain thermoreflectance (TDTR) and compared with literature values. Furthermore, finite element analysis (FEA) was conducted on a back-gated indium-tin-oxide (ITO) device to evaluate the thermal performance of AlN. The combination of experimental measurements and thermal simulations provides a comprehensive understanding of the thermal management potential of AlN films in practical BEOL integration.

## 2. Sample Preparation and Experiments

### 2.1 Sample Preparation

The AlN thin films were deposited on commercially available Si, SiO, SiN and $Al_2O_3$ (sapphire) substrates at 400 °C using magnetron sputtering. Two different AlN thicknesses (~600 nm and ~1200 nm) were obtained for each substrate, resulting in 8 combinations in total as listed in Table I. An ~80-nm Al transducer layer was later deposited on top of the samples for TDTR measurements using an electron beam evaporation system.

Table I. AlN thicknesses and substrates of the 8 samples in this study. The SiO/SiN substrates are thermally thick SiO/SiN layers grown on Si substrates.

| Thickness | Substrate | | | |
|---|---|---|---|---|
| | Si<111> | SiO | SiN | $Al_2O_3$ |
| ~600 nm | S1 | S2 | S3 | S4 |
| ~1200 nm | S5 | S6 | S7 | S8 |

### 2.2 Structural and Thermal Characterization

Following AlN deposition, XRD and STEM were used to examine the quality of the samples. The STEM image also verifies the thicknesses of AlN films measured by picosecond acoustics, which is a pump-probe method for thickness measurement of thin films[24]. Picosecond acoustics leverages the thermal expansion induced by laser pulses to generate acoustic waves that are reflected back from material interfaces. This system is an inherent part of the TDTR system discussed below. For thermal characterizations, the thermal conductivity of AlN thin films was measured using the TDTR method. TDTR is a solid pump-probe technique that extracts the thermal properties of layered samples, which leverages a femtosecond laser source and has been detailed in previous literature[9], [25]. In the two-tint TDTR setup used in this research, the laser beam is first generated by a Mai Tai sapphire:Ti laser source operating at 80 MHz with a center wavelength of 785 nm and then spilt into pump and probe beams. The pump beam is further modulated at 5.02 MHz through an electro-optic modulator (EOM). The radii of the pump and probe beam spots are both 6.0 μm. It should be noted, however, that the TDTR signal may not be sensitive to the TC of excessively thin layers due to their limited contribution to the overall thermal resistance of the sample. Therefore, the thicknesses of AlN films in this research were taken large enough to ensure good sensitivity of their TC. On the other hand, we have also noticed anomalously high TC reported for <300-nm AlN films in literature which even exceeds the theoretical upper limit of thin-film AlN[26]. Such results might be better attributed to the error during the TDTR fitting process and could be improved by the dual-frequency TDTR approach[27].

## 3. Results and Discussion

### 3.1 Structural and Thermal Properties

Fig. 1 shows the high-resolution XRD rocking curves of the 8 samples in this study. The full width at half maximum (FWHM) values of 1200-nm films are generally smaller than those of 600-nm films, indicating higher crystal quality for regions farther from the AlN/substrate interface[28]. The smallest peak widths were observed in AlN grown on sapphire, which helps explain the largest TC values found in these samples.

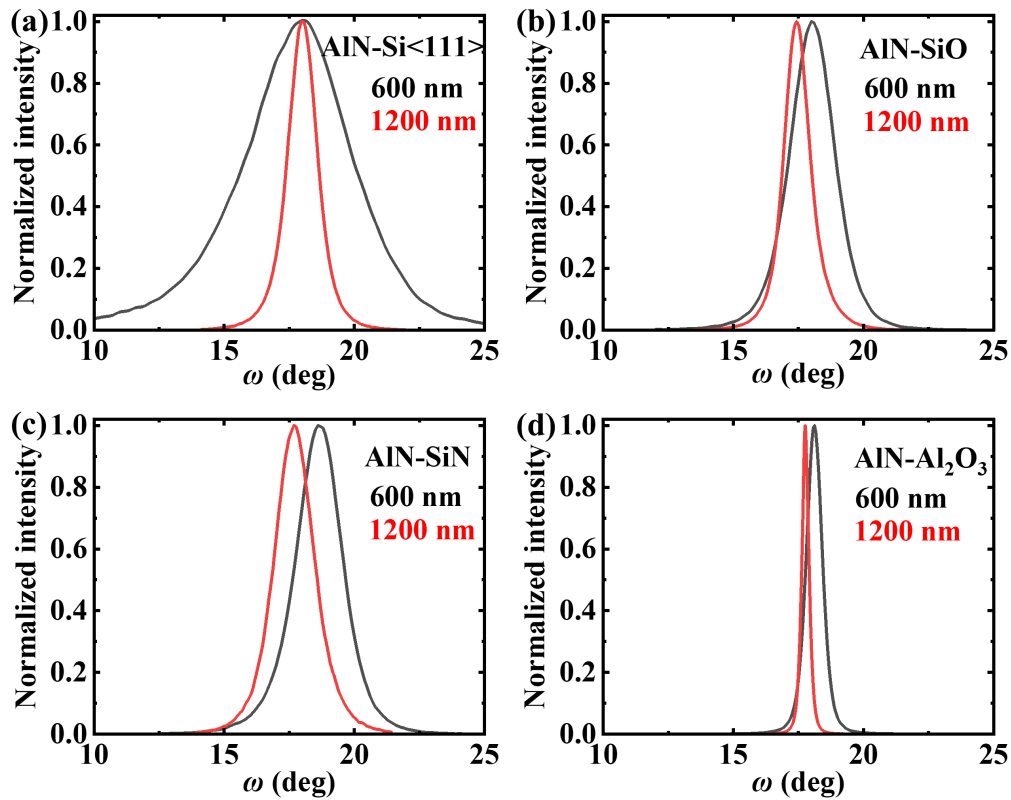

Fig. 1. XRD rocking curves of the samples. Black lines indicate 600-nm samples, while red lines indicate 1200-nm ones. The shifts in peaks among different samples are possibly caused by residual stress during AlN deposition. (a) AlN on Si<111> (600-nm AlN FWHM = 4.8°, 1200-nm AlN FWHM = 1.3°). (b) AlN on SiO (600-nm AlN FWHM = 2.0°, 1200-

nm FWHM = 1.2°). (c) AlN on SiN (600-nm AlN FWHM = 2.1°, 1200-nm AlN FWHM = 1.8°). (d) AlN on sapphire (600-nm AlN FWHM = 0.7°, 1200-nm AlN FWHM = 0.3°).

Fig. 2(a-c) shows the high-angle annular dark-field STEM (HAADF-STEM) images of 1.2-μm AlN grown on $Al_2O_3$ taken by ThermoFisher Talos F200X, which clearly indicate the (0002) growth direction, polycrystalline structure and columnar grains. Prior to TDTR measurements, the thicknesses of AlN films were also verified by picosecond acoustics in the TDTR system (Fig. 2(d-e)).

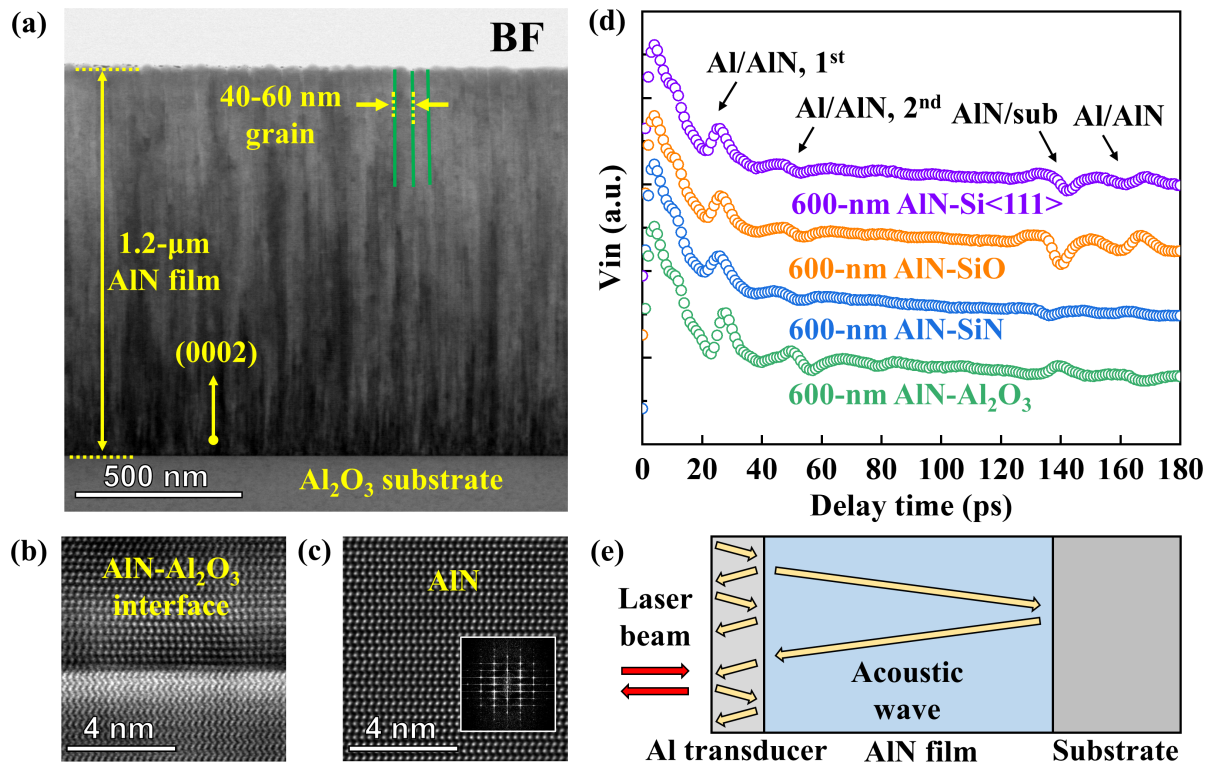

Fig. 2. (a) Cross-sectional HAADF-STEM images of the AlN-sapphire sample. Columnar AlN structures are marked with green lines indicating grain boundaries. The grain size is estimated to be 40-60 nm. (b) High-resolution image at the AlN-sapphire

interface showing no amorphous interlayer. (c) High-resolution image and FFT result (inset) in the AlN region which indicates the (0002) growth direction. (d) Picosecond acoustics measurements on 600-nm AlN samples. The curves are shifted along the Y axis for ease of distinguishing. (e) Schematic of the picosecond acoustics measurement.

The TDTR signal was collected and fitted to an analytic heat transfer model to extract the TC of AlN at room temperature (Fig. 3(a-b)), with the highest TC value occurring in the 1200-nm AlN sample deposited on an $Al_2O_3$ substrate (detailed values are reported in Table SI). Even though the TC values are lower than the theoretical upper limit for defect-free crystal, they remain relatively high for heat spreading purposes. This may be attributed to the high-quality growth of AlN polycrystalline films, which yields large grain size, uniform growth orientation and thus reduced boundary scattering. Meanwhile, the TCs of 1200-nm AlN samples on Si and SiO substrates were further measured over 300-550 K (Fig. 3(c)). Compared to bulk material, the TC of thin-film AlN is less sensitive to temperature due to the existence of phonon boundary scattering, since its scattering rate changes more slowly with temperature than that of phonon-phonon scattering which is the dominant scattering mechanism in defect-free bulk AlN.

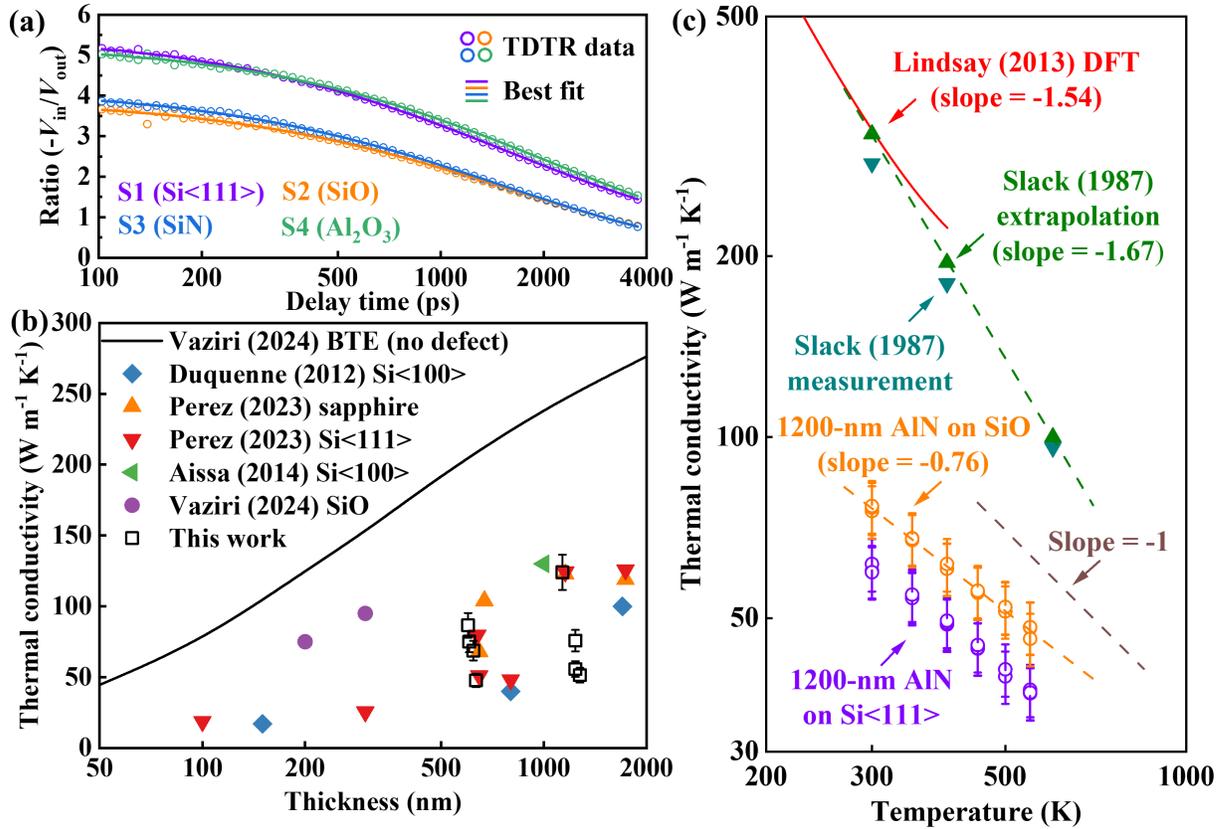

Fig. 3. (a) TDTR data and best fit curves for 600-nm AlN samples. (b) Measured and literature TC values of AlN thin film samples deposited at BEOL-compatible low temperature[31], [32], [33], [34]. (c) TC values of 1200-nm AlN thin films grown on Si<111> (red circles) and SiO (black circles) measured over 300-550 K, compared with literature data of bulk AlN samples from Slack et al.[11] (measurement, brown dots; extrapolation, purple dots) and Lindsay et al.[26] (DFT, blue line). Refs [28], [29], and [30] report bulk TC values that have similar trends and are not shown in the figure. The error bar is estimated at 10%, a typical value for TDTR measurements.

Fig. S1 shows the sensitivity curves for parameters of interest during TDTR fitting.

Although the sensitivity of AlN TC remains reasonably high in all samples, the sensitivity of the TBC at AlN/SiO and AlN/SiN interfaces is low due to the large thermal resistance of the substrates. These TBC values are fixed in the fitting process at 100 MW/m$^2$-K, which is a typical value for dielectric-dielectric interfaces.

It should be noted here that the TC value measured here is the cross-plane TC, as a normal TDTR configuration could be insensitive to in-plane heat transfer[35]. In-plane TC is as important as cross-plane TC for heat spreaders, which has been measured in previous literature using steady-state techniques[31], [36].

## 3.2 FEA Modeling and Analysis

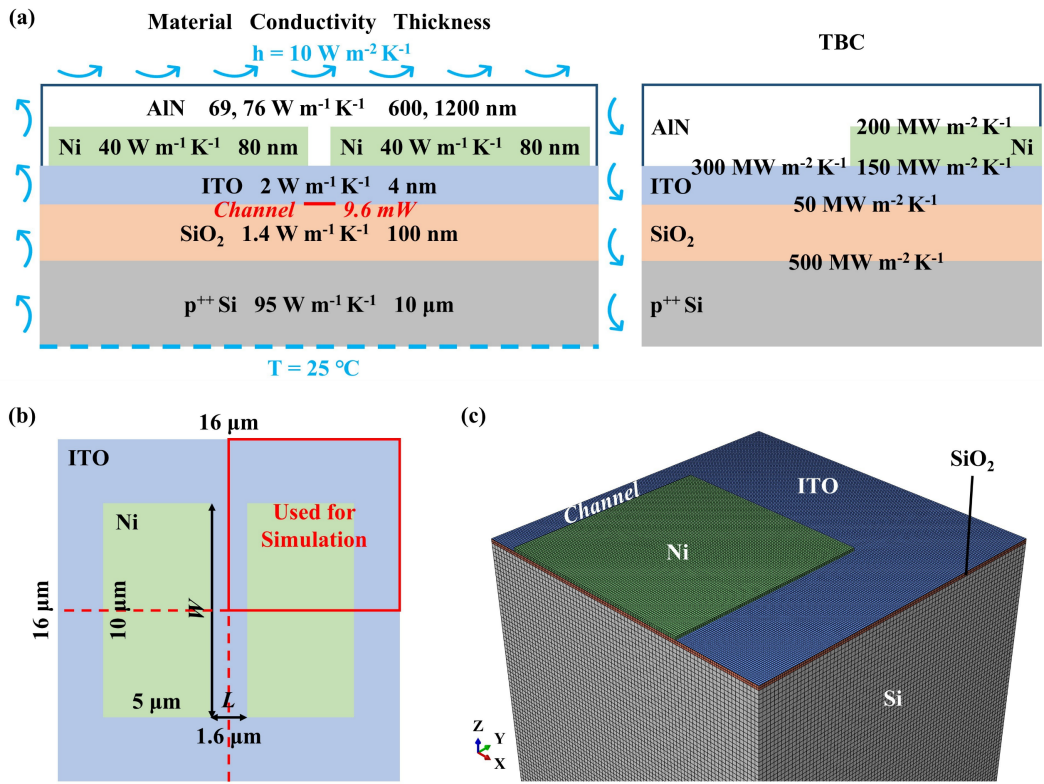

Fig. 4. Finite element model of the back-gated ITO device with 9.6 mW power uniformly dissipated across the channel. Device schematics of (a) the cross section and (b) the top view, showing the component materials, boundary conditions, geometric dimensions, TC values, and TBC values. (c) Meshing of the model.

Finite element analysis (FEA) was conducted in ABAQUS to examine the thermal management performance of AlN in a back-gated ITO device. Fig. 4(a) and Fig. 4(b) schematically show the cross section and top view of the model. The back-gated ITO device incorporates Ni contacts and ITO thin film deposited on $SiO_2$ and back-gate Si substrate. As shown in Fig. 4(a), a total power of 9.6 mW was uniformly dissipated across the channel.

Natural convection was applied to the top and side surfaces of the device, and the bottom temperature was maintained at 25 °C. The initial temperature of the device was set at 25 °C.

Fig. 4(a) also summarizes the isotropic TCs of all components and TBCs between different materials. Based on the experiment results, TC values of 68.7 W m$^{-1}$ K$^{-1}$ and 75.8 W m$^{-1}$ K$^{-1}$ were employed for 600- and 1200-nm AlN, respectively. As shown in Fig. 4(b) and Fig. 4(c), a quarter model was constructed for simulation considering the symmetry, with a verified mesh consisting of 736,800 elements. In parametric studies, the model was modified to investigate the effects of channel length, TBC of AlN-ITO, AlN thickness and coverage configuration.

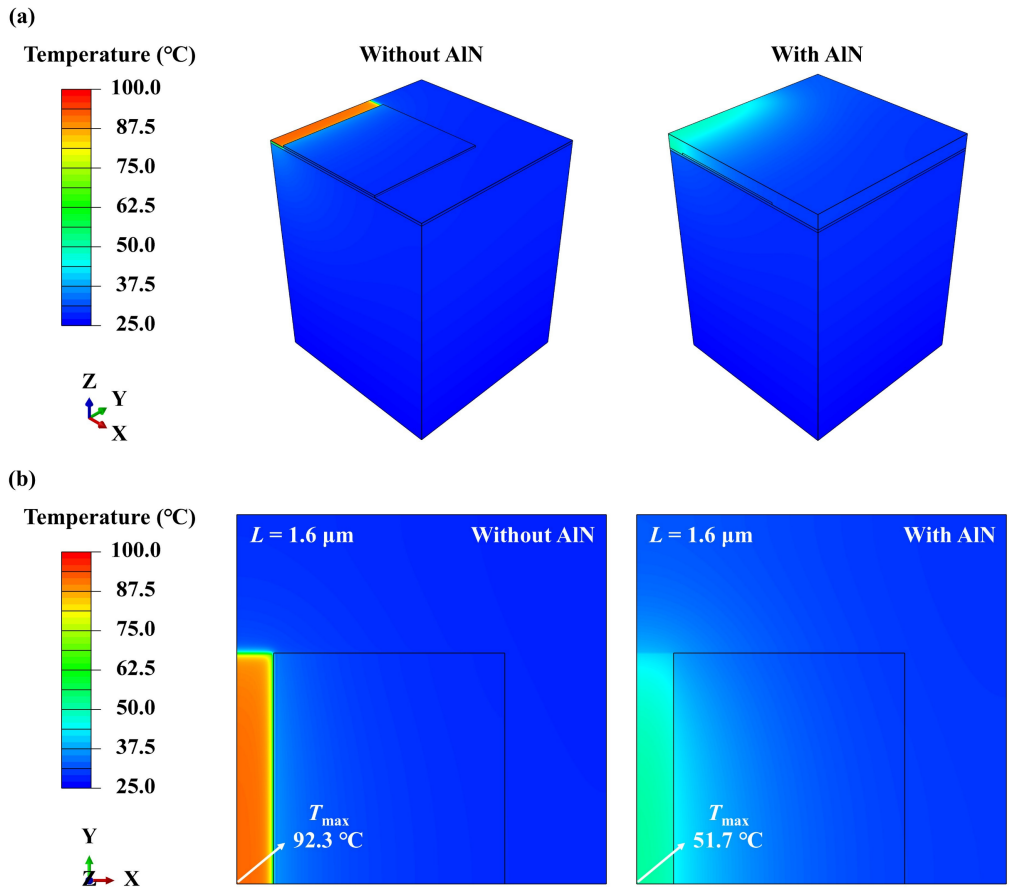

Fig. 5. Simulated temperature distributions of the ITO device with and without AlN (a) across the entire model and (b) along the Ni/ITO interface.

Fig. 5 compares the steady-state temperature distribution of the back-gated ITO device with and without AlN. As shown in Fig. 5(a), in the absence of AlN, the channel region exhibits a pronounced temperature rise. Heat is primarily confined to the active area, indicating limited lateral heat spreading, which can be attributed to the low thermal conductivity of ITO (2 W m$^{-1}$ K$^{-1}$). In contrast, with the introduction of the AlN layer, heat generated in the channel is conducted into the overlying AlN layer, thereby enhancing lateral heat spreading due to the high thermal conductivity of AlN (68.7 W m$^{-1}$ K$^{-1}$). As a result, the peak device temperature exhibits a ~44% reduction, decreasing from 92.3 °C to 51.7 °C, as shown in Fig. 5(b).

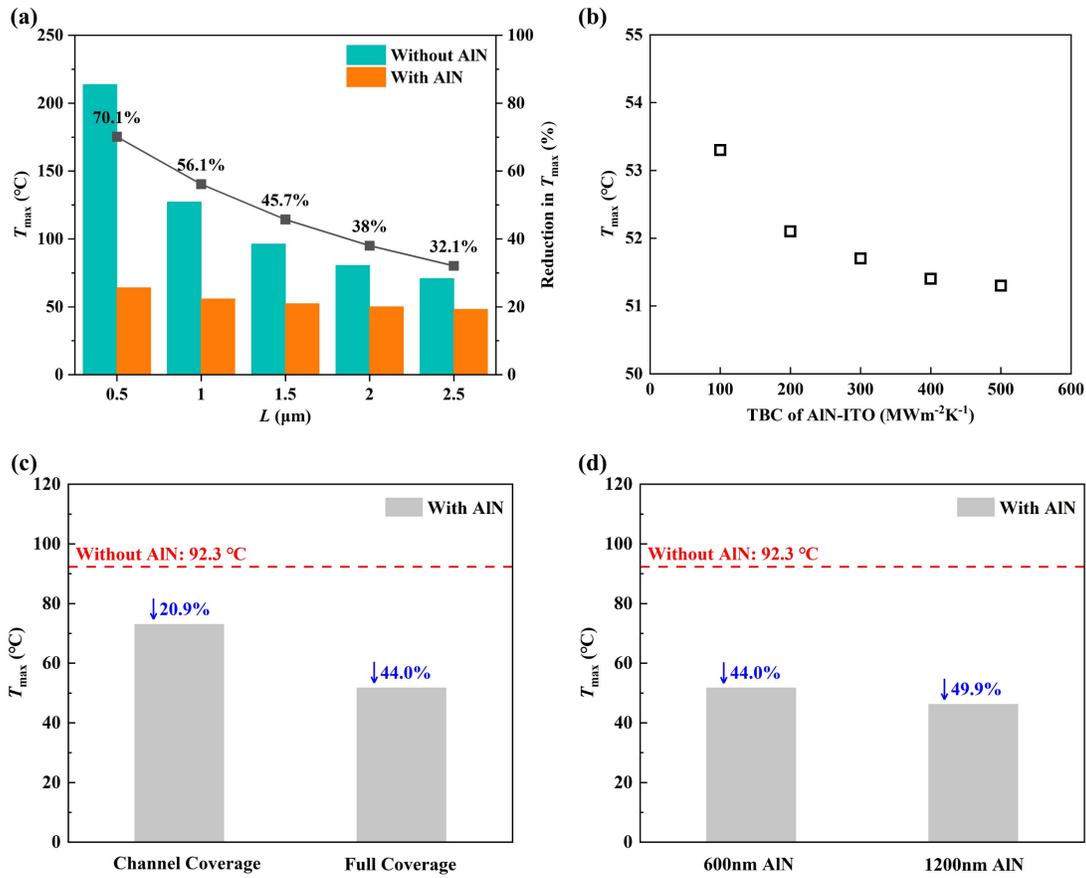

Fig. 6. Peak device temperature of 9.6 mW input power under different (a) channel length, (b) AlN–ITO thermal boundary conductance, (c) AlN coverage configuration, and (d) AlN thickness.

Fig. 6 (a) shows that AlN coverage reduces the peak device temperature across all investigated channel lengths. As the channel length decreases, the reduction in peak temperature becomes more pronounced. According to the temperature distributions shown in Fig. S2, device with a shorter channel exhibits more severe heat accumulation due to the increased heat flux and the low thermal conductivity of ITO, which makes lateral heat spreading increasingly difficult. AlN coverage directly addresses this limitation by enhancing lateral heat spreading, thereby resulting in a particularly pronounced temperature reduction for shorter channels. Notably, AlN coverage on the device with a 0.5-μm channel reduces the peak temperature from 213.7 °C to 63.9 °C, which significantly improves the thermal reliability.

Beyond geometric scaling, interfacial heat transfer plays a critical role in cooling performance, as illustrated in Fig. 6(b). As the AlN-ITO TBC increases, heat transfer from the channel into the AlN layer is enhanced, leading to a reduction in the peak device temperature. The improved interfacial heat transfer allows the AlN layer to promote lateral heat spreading more effectively, as shown in Fig. S3. Moreover, at high TBC values, further increase in TBC produces only a marginal reduction in the peak device temperature. The diminishing rate of temperature reduction can be attributed to the increasing dominance of bulk thermal resistance over interfacial resistance as TBC increases.

To further investigate the heat dissipation performance of AlN, two configurations of AlN coverage were evaluated, as shown in Fig. 6(c). Channel coverage reduces the peak device temperature by 20.9%, whereas full coverage results in a more pronounced reduction of 44.0%. As shown in Fig. S4, for channel coverage, the AlN layer provides a thermal path from the channel to the Ni contacts where heat is laterally dissipated. In contrast, for full coverage, the AlN layer functions not only as a thermal path but also as a heat spreader, allowing heat to be distributed over a larger area. Consequently, channel coverage offers limited thermal improvement compared with full coverage.

Apart from coverage configuration, the influence of AlN thickness is also presented in Fig. 6(d). Compared with the device without AlN, the 600- and 1200-nm AlN layers reduce the peak device temperature by 44.0% and 49.9%, respectively. The 1200-nm AlN layer exhibits a higher thermal conductivity of 75.8 W m$^{-1}$ K$^{-1}$, which enhances lateral heat spreading and results in a slightly greater temperature reduction. Notably, the difference in peak temperature between the two thicknesses is smaller than that between the two coverage configurations, as illustrated in Fig. S4 and Fig. S5.

## Conclusion

In summary, this work presents experimental and computational evidence for the potential of AlN to be used as efficient heat spreaders in microelectronics. With the combination of structural characterization (XRD and STEM) and thermal characterization (TDTR) methods, we found that 600- and 1200-nm AlN thin films can be sputter-deposited at

BEOL-compatible low temperature (400 °C) on various substrates like Si, SiO, SiN and Al$_2$O$_3$ while retaining relatively high TC (> 45 W m$^{-1}$ K$^{-1}$). The TCs of AlN samples vary with the substrate material yet remain consistently satisfying for heat spreading purposes. By examining the temperature distribution of AlN capped on a back-gated ITO device, we further verified the heat spreading performance of AlN thin films. These findings provide a perspective on solving the pressing thermal issues in 3D-IC and could aid in further research into the process optimization and integration of AlN material.

## Acknowledgments

The authors acknowledge the financial support from the National Key Research and Development Program of China (Grant No. 2024YFA1207900) and the National Natural Science Foundation of China (NSFC) (Grant Nos. 62574007, T2550270).

## Competing Interests

The authors declare no competing interest.

## Data Availability

The data that support the findings of this study are available from the corresponding authors upon reasonable request.